\documentclass[12pt]{iopart}
\usepackage{iopams,amsopn}
\usepackage[square,sort&compress,numbers]{natbib}
\usepackage[dvips]{graphicx}
\DeclareMathOperator{\sech}{sech}
\DeclareMathOperator{\csch}{csch}

\begin{document}

\paper{Localization in the ground level state of a triple quantum well}

\author{R. Mu\~{n}oz-Vega$^1$, L. Diago-Cisneros$^2$, J.-J. Flores-Godoy$^3$ and G. Fern\'andez-Anaya$^3$}
\address{$^1$ Universidad Aut\'{o}noma de la Ciudad de M\'{e}xico, Centro Hist\'{o}rico, Fray Servando Teresa de Mier 92, Col. Centro, Del. Cuauht\'{e}moc, M\'{e}xico D. F.,  06080, M\'exico}
\ead{rodrigo.munoz@uacm.edu.mx}
\address{$^2$ Facultad de F\'{i}sica, Universidad de la Habana, C. P. 10400, La Habana, Cuba}
\ead{ldiago@fisica.uh.cu}
\address{$^3$ Departamento de F\'isica y Matem\'aticas, Universidad Iberoamericana, Prol. Paseo de la Reforma 880, Col. Lomas de Santa Fe, Del. A. Obreg\'on, M\'exico, D. F.  01219, M\'exico}
\eads{\mailto{job.flores@ibero.mx}, \mailto{guillermo.fernandez@ibero.mx}}


\begin{abstract}
A model is presented, consisting of a single structureless particle on the line subject to a potential with three minima, with an exactly soluble ground level. In this model the ground level probability density becomes more sensitive to the global shape of the potential as the distance between the minima increases, so that for big enough distances small variations in the potential bring a qualitative change in the probability density, taking it from a unimodal, localized, distribution, to a bimodal one. We conjecture that this effect, of which we have not found any precedent in the literature,  may be relevant in the design and characterization. of mesoscopic devices such as triple quantum well systems.
\end{abstract}
\date{\today}

\maketitle
\section{Introduction}

Triple quantum well systems are of current technological interest, with applications in lasing, optical detection and modulation, and other fields.\cite{Ghoshetal,Zhao,Sun,Han,PING1} Related to this, Multiple Quantum Wells (MQW) are now being applied in the design of alternative solar cells.\cite{Barnham} This latter devices might extend the spectral response of conventional solar cells, and thus yield an increased photocurrent without  an increase in open circuit voltage degradation.\cite{Harrison}  A growing demand for MQW based solar cells constitutes by itself a sufficient motivation for the study of the electronic properties of such systems, and thus for the sudy of their confining potentials.

Also, mesoscopic semiconductor heterostructures and other mesoscopic systems exhibit a rich phenomenology that mingles quantum and classical aspects,\cite{Dykman} enticing the interest of  theoreticians concerned with fundamental questions such as the limits of quantum theory.\cite{Leggett2002}

In the present paper we present a mathematical procedure that produces (admittedly
simple) models of MQWs with known ground eigenvalues and eigenfunctions, paying special attention to a family triple well potentials. We find that this latter family behaves in a possibly counterintuitive manner, of which we have not found precedent in the literature. We believe that this may be of interest to both applied researchers and their more theoretically minded colleagues.   

Suppose we approximate, in a semi-classical fashion, the ground level, $\Phi(x)$, of a one-dimensional potential $W(x)$ with $N$ wells as the  coherent superposition of $N$ states, $\phi_{1}$, $\phi_{2}\ldots\phi_{N}$, each localized around a different well:
\begin{equation}
\Phi (x)\approx\sum\limits_{j=1}^{N} c_{j}\phi_{j}(x),
\end{equation}
with coefficients $c_{j}$ such that both $\Phi$ and the $\phi_{j}$ are normalized. If the distances, $a_{1},\ldots ,a_{N-1},$ between adjacent wells are all increased while changing neither the depth nor the width of each well, then the localization of the $\phi_{j}$ implies that the overlap integrals
\begin{equation}
\int \phi_{j}^{*}(x)\phi_{k}(x)dx, \; j\neq k
\end{equation}
vanish as $a\rightarrow\infty$, so that the $\phi_{j}$ become linearly independent and the ground level becomes $N$-fold degenerate.

For $a<\infty$ a potential bounded from below can never be truly degenerate.\cite{Landau} Instead, when the separation between wells is big enough a low-lying effective $N$-level system appears, exhibiting the oscillatory phenomena associated which such systems. Yet, the frequencies associated with this oscillations (frequencies proportional to the differences in the energies in the $N$-level system) may be small enough as to make the oscillations unobservable, due to dissipation or some other mechanism. In this later case there is probably no way to retrieve the values of the coefficients $c_{j}$ from experimental results, so that under this circumstances the $c_{j}$  must be considered as indeterminate quantities (save from a global normalization condition). On the other hand, for any given potential $W:\mathbb{R}\rightarrow\mathbb{R}$, if the potential is known beforehand, the $c_{j}$ cannot be chosen at whim, but are to be calculated, \emph{viz}, by minimizing the ground level through a variational procedure.

In the following pages we construct a family of potentials $V_{\lambda, a}$ with an exact expression for its corresponding ground eigenstates $\Psi^{(\lambda, a)}$ in the form
\begin{equation}
\Psi^{(\lambda, a)}(x)=\sum\limits_{j=1}^{3}c_{j}^{(\lambda, a)}\psi (x+a_{j}),
\end{equation}
where $\psi(x)$ is a normalized function, a stationary solution of some other potential. In this model the (real valued) $a_{j}$ parameters are approximations to the positions of the three minima of the $V_{\lambda, a}$ potential, approximations that become more accurate when the distances, $\left\vert a_{j}-a_{k}\right\vert$ ($j\neq k$), are increased simultaneously. There is an extra parameter, $\lambda$, (also real) that affects the overall shape of the potential. In the cases when all distances, $\left\vert a_{j}-a_{k}\right\vert$, are above a certain threshold, it is observed that the quotients $V_{\lambda , a}(a_{j}) / V_{\lambda , a}(a_{k})$, which give an approximation of the relative depths of the wells, become practically independent of the value of $\lambda$, while the quotients $\left\vert c_{j}^{(\lambda , a)}/c_{k}^{(\lambda , a)}\right\vert^{2}$, which give approximations to the relative height of the probability density peaks in different wells, remain quite sensitive to value of $\lambda$, especially when all distances are above the threshold. Thus, when the distances between the minima of the potential become sufficiently big, the ground level probability density becomes exquisitely sensitive to the relative depth of the wells of a $V_{\lambda , a}$ potential.

If the experimental determination of the $c_{j}$ coefficients depends on the condition that the low-lying energy levels can be resolved, while, on the other hand, it is known that the mentioned  effect will appear only when the distances between wells surpass a certain threshold, then these contradictory demands would surely pose a challenge to any experimentalist willing to take this paper in consideration.  A challenge that, as is argued in the following pages, may be insurmountable. The present paper is dedicated to discuss the consequences of this apparent paradox: that of an effect that, while expected by the theory, is at the same time predicted to be unobservable in foreseeable realizations.

Submicron semiconductor heterostructures of low dimensionality, including quantum wells, quantum dots and related systems, have now been synthesized  and studied for over thirty years. Coupling effects between adjacent wells were reported as early as 1975, \cite{Dingle} and have been studied ever since. Phenomena that can be arguably be related to this coupling, such as single-electron oscillations in the tunneling across junctions \cite{vanderZant} and resonance-like oscillations in the electrical conductance at mili-Kelvin temperatures \cite{Meirav} have also been known to exist for quite some time. Moreover, the operation of such devices has steadily being extended to the single electron regime \cite{Meirav,Geerligs,Waughetal,Ciorga,Elzerman}. Thus, we gather that  the experimental realization of a system similar to our model is feasible.

The rest of this paper is structured as follows: In Section 2 we present a procedure for the construction of potentials with $N$ wells and exactly solvable ground levels, and  discuss some of the particulars of the potentials thus obtained. We then go on to focus on the construction of a symmetric potential with three minima, in Section 3. The aforementioned sensitivity-growing-with-distance effect shown by the example of Section 3 is discussed in Section 4. Then, some reference values for the experimental observation of our results are laid out  in Section 5. Finally, Section 6 is dedicated to discuss the possible significance of our results. Some tentative conclusion are advanced in this last Section.
%
%
%
%
\section{A procedure for the construction of potentials with known ground eigenfunctions}\label{sec:2}
Consider an adimensional version of a Hamiltonian $H$, of the form:
\begin{equation}\label{initial.H}
H=-\frac{d^{2}}{dx^{2}}+V(x)\quad\textrm{ , }\quad V:\mathbb{R}\rightarrow\mathbb{R}
\end{equation}
in a system of units such that $\hbar^{2}/2m=1$, with $x$ an adimensional coordinate and $V(x)$ a non-singular potential  with a bounded ground energy level $E_{0}$ with known corresponding eigenfunction  $\psi_{0}$.  Then, for each finite $N$-tuple of real numbers $\Lambda=(\lambda_{1},\lambda_{2},\ldots,\lambda_{N})$ such that
\begin{equation}\label{convex.def}
\lambda_{k} > 0 \quad k=1,2,\ldots , N
\end{equation}
and that
\begin{equation}\label{convex.def.II}
\sum\limits_{n=1}^{N}\lambda_{n}=1,
\end{equation}
and each $N$-tuple of real numbers $A=( a_{1}, a_{2},\ldots , a_{N})$, a  Hamiltonian $H_{\Lambda, A}$ can be constructed that shares its ground energy level, $E_{0}$ with $H$ and that has as ground eigensolution
\begin{equation}\label{ground}
\Psi_{0}^{ (\Lambda, A)}(x)= \alpha_{\Lambda, A} \sum_{n=1}^{N}\lambda_{n}\psi_{0}(x+a_{n}),
\end{equation}
with $\alpha_{\lambda, A}$ a constant.

Indeed, as $H$ is one-dimensional, an arbitrary global phase can be chosen as to make $\psi_{0}$ real-valued, \emph{i.e.,} we can always take $\psi_{0}:\mathbb{R}\rightarrow\mathbb{R}$. Furthermore, as $\psi_{0}$ represents a ground state, it is then free of nodes, and can thus be taken without loss of generality as positive definite, \emph{i.e.,} we can consider that $\psi_{0}:\mathbb{R}\rightarrow\mathbb{R}^{+}$. Consequently, the function $\Xi_{\Lambda , A}:\mathbb{R}\rightarrow\mathbb{R}^{+}$, defined through
\begin{equation}\label{xi.def}
\Xi_{\Lambda, A}(x)=\sum\limits_{n=1}^{N}\lambda_{n}\psi_{0}(x+a_{n})
\end{equation}
will also be node-free, and the potential $V_{\Lambda, A}:\mathbb{R}\rightarrow\mathbb{R}$ given by
\begin{equation}\label{V.def}
V_{\Lambda,A}(x)=\frac{\sum\limits_{n=1}^{N}\lambda_{n}V(x+a_{n})\psi_{0}(x+a_{n})}{\Xi_{\Lambda,A}(x)}
\end{equation}
will be non-singular, as $V$ has been chosen non-singular. The ground level eigenstate of Hamiltonian
\begin{equation}\label{H.def}
H_{\Lambda , A}=-\frac{d^{2}}{dx^{2}}+V_{\Lambda , A}(x)
\end{equation}
is represented by the function (\ref{ground}) as can be verified by plugging (\ref{V.def}) in (\ref{H.def}) and applying the result to (\ref{ground}).  The normalization factor $\alpha_{\Lambda , A}$, given by
\begin{equation}\label{A.def}
\alpha_{\Lambda , A}=\Bigg(\int_{-\infty}^{\infty}\Big[\Xi_{\Lambda , A}(x)\Big]^{2}dx\Bigg)^{-1/2}\textrm{,}
\end{equation}
is a finite, strictly positive, real number for each $A\in\mathbb{R}^{N}$. Indeed, by inserting definition (\ref{xi.def}) in (\ref{A.def}) we obtain the relation
\begin{equation}\label{norm.rel}
\alpha_{\Lambda , A}=\Bigg(\sum\limits_{n=0}^{N}\lambda_{n}^{2}+\sum_{k\neq l}^{N}\lambda_{k}\lambda_{l}\mathcal{O}_{k,l}^{(A)}\Bigg)^{-1/2}.
\end{equation}
where $\mathcal{O}_{k,l}^{(A)}$ stands for the overlap integral
\begin{equation}\label{overlap.i}
\mathcal{O}_{k, l}^{(A)}=\int_{-\infty}^{\infty}\psi_{0}(x+a_{k})\psi_{0}(x+a_{l})dx\ .
\end{equation}
Thus
\begin{equation}
1<\alpha_{\Lambda , A}<\Big(\sum\limits_{n=0}^{N}\lambda_{n}^{2}\Big)^{-1/2} \ .
\end{equation}
In this way we have proven our assertion: that given a one-dimensional Hamiltonian $H$ with a bounded ground eigenstate and $N$-tuples $\Lambda$ and $A$ complying with (\ref{convex.def}) and (\ref{convex.def.II}), a Hamiltonian $H_{\Lambda, A}$ can be constructed that shares its ground level with $H$ and has (\ref{ground}) as ground eigensolution.

From definitions (\ref{xi.def}) and (\ref{V.def}) we have that for any given $ x\in\mathbb{R}$
\begin{equation}
\min\limits_{a\in A} V(x+a) \leq V_{\Lambda , A}(x) \leq \max\limits_{a\in A} V(x+a)
\end{equation}
so that, if the initial potential $V(x)$ is bounded from below by  a real constant $V_{L}$, that is, if
\begin{equation}
V_{L}\leq V(x) \quad \forall x\in\mathbb{R}
\end{equation}
then
\begin{equation}
V_{L}\leq V_{\Lambda , A}(x)\quad \forall x\in\mathbb{R}
\end{equation}
for any $N$-tuples $\Lambda$ and $A$ we choose. An analogous property holds for the upper bounds (if any) of $V(x)$.

We now turn our attention to the ground level probability density $\rho_{\Lambda, A}(x)=\vert\Psi_{0}(x;A, \Lambda)\vert^{2}$ which can be written in the suggestive form
\begin{equation}\label{quant.dist}
\rho_{\Lambda , A}(x)=\iota_{\Lambda , A}(x)+\sum\limits_{n=0}^{N}W_{n}^{(\Lambda , A)}\rho(x+a_{n})
\end{equation}
with the use of equations (\ref{ground}), and (\ref{norm.rel}). Here, $\rho(x)=\vert\psi_{0}(x)\vert^{2}$ is just the probability distribution of the ground eigenstate of the original Hamiltonian $H$, the $W^{(\Lambda , A)}_n$ are the `weighting factors'
\begin{equation}\label{w.factors}
W_{n}^{(\Lambda , A)}=\alpha_{\Lambda, A}^{2}\lambda_{n}^{2}
\end{equation}
and `overlap term' $\iota_{\Lambda , A}(x)$ is as given by
\begin{equation}\label{int.def}
\iota_{\Lambda , A}(x)=\alpha_{\Lambda , A}^{2}\sum_{k\neq l}^{N}\lambda_{k}\lambda_{l}\psi_{0}(x+a_{k})\psi_{0}(x+a_{l}).
\end{equation}
Notice that $\rho(x+a_{n})$ is the probability density for the ground state of Hamiltonian:
\begin{equation}\label{neg.def}
H_{n}=-\frac{d^{2}}{dx^{2}}+V(x+a_{n}), \quad n=1,2, \ldots , N .
\end{equation}
If the `overlap term' was to to be negligible in expression (\ref{quant.dist}), along with all overlap integrals, \emph{i.e.,} if
\begin{equation}\label{distinct.cond}
\iota_{\Lambda, A}(x)\ll \sum\limits_{n=1}^{N}W_{n}^{(\Lambda , A)}\rho(x+a_{k}) \quad \forall a\in A, \forall x\in\mathbb{R},
\end{equation}
and
\begin{equation}\label{null.overlap}
\sum_{k\neq l}^{N}\lambda_{k}\lambda_{l}\mathcal{O}_{k,l}^{(A)}\ll \sum\limits_{n=1}^{N}\lambda_{n}^{2}\quad ,
\end{equation}
then the probability density $\rho_{\Lambda , a}$ would approach the distribution
\begin{equation}\label{class.dist}
\rho_{\Lambda , A}^{\mathrm{class}}(x)=\sum\limits_{n=1}^{N}\Bigg(\lambda_{n}^{2} \Bigg [ \sum\limits_{k=1}^{N}\lambda_{k}^{2}\Bigg ]^{-1}\Bigg)\rho(x+a_{n}).
\end{equation}

Let us now see what can be said about the excited energy levels of the $H_{\Lambda , A}$ . To this end, consider an $N$-tuple $\beta\in\mathbb{R}^{N}$ and a normalized trial function $F_{\beta , A}:\mathbb{R}\rightarrow\mathbb{R}$ which is just a linear combination of the $\psi_{0}(x+a_{k})$, that is
\begin{equation}\label{f.trail}
F_{\beta A}(x)=\sum\limits_{k=1}^{N}\beta_{k}\psi_{0}(x+a_{k}) \ ,
\end{equation}
with a normalization condition
\begin{equation}\label{n.oder.norm.cond}
\sum\limits_{k=1}^{N}\beta_{k}^{2}+\sum\limits_{n\neq k}^{N,N}\beta_{n}\beta_{k}\mathcal{O}_{n,k}^{(A)}=1
\end{equation}
written in terms of the overlap integrals of (\ref{overlap.i}).

By inserting (\ref{H.def}) in  front of (\ref{f.trail}), and taking definition (\ref{V.def}) into account, one obtains
\begin{equation}\label{Hf}
H_{\Lambda, A}F_{\beta, A}(x)=\sum\limits_{k=1}^{N}\beta_{k}\Big[E_{0}-V(x+a_{k})+V_{\Lambda , A}(x)\Big]\psi_{0}(x+a_{k})
\end{equation}
and from (\ref{Hf}) it is immediate that
\begin{eqnarray}
\fl \int_{-\infty}^{\infty}F_{\beta, A}(x) H_{\Lambda , A} F_{\beta , A}(x)\ dx = E_{0} + \nonumber \\
\sum\limits_{n=1}^{N}\beta_{n}^{2}\Bigg(\mathcal{V}_{n,n}^{(\Lambda , A)}-\langle \ V\rangle\Bigg)+\sum\limits_{k\neq n}^{N,N}\beta_{k}\beta_{n}\Bigg(\mathcal{V}_{n,k}^{(\Lambda , A)}-\mathcal{U}_{k,n}^{(A)}\Bigg)
\label{integral.f.relation}
\end{eqnarray}
where $\langle\ V\rangle$, $\mathcal{V}_{k,n}^{(\Lambda , A)}$ and $\mathcal{U}_{k,n}^{(A)}$ stand, respectively, for
\begin{equation}\label{expect.v}
\langle\ V\rangle=\int_{-\infty}^{\infty}\psi_{0}(x)V(x)\psi_{0}(x)\ dx \quad ,
\end{equation}
\begin{equation}\label{nother.bee}
\mathcal{V}_{k,n}^{(\Lambda , A)}=\int_{-\infty}^{\infty}\psi_{0}(x+a_{n})V_{\Lambda , A}(x)\psi_{0}(x+a_{k})\ dx
\end{equation}
and
\begin{equation}\label{bumble.bee}
\mathcal{U}_{k,n}^{(A)}=\int_{-\infty}^{\infty}\psi_{0}(x+a_{n})V(x+a_{k})\psi_{0}(x+a_{k})\ dx\ .
\end{equation}
In this paper we shall focus in the case when $V(x)$ has both an upper bound, $V_{U}$, and a lower bound, $V_{L}$, so that, because of (\ref{integral.f.relation}) the relation
\begin{eqnarray}
\fl \int_{-\infty}^{\infty}F_{\beta, A}(x) H_{\Lambda , A} F_{\beta , A}(x)\ dx\leq E_{0}+ \nonumber \\
\sum\limits_{n=1}^{N}\beta_{n}^{2}\Bigg(\mathcal{V}_{n,n}^{(\Lambda , A)} -\langle \ V\rangle \Bigg)+\Bigg(V_{U}-V_{L}\Bigg)\sum\limits_{k\neq n}^{N,N}\beta_{k}\beta_{n}\mathcal{O}_{k,l}^{(A)}\label{alice.in.chains}
\end{eqnarray}
stands.

In principle, upper bounds
\begin{equation}\label{bounds.momma}
E_{n}^{(\Lambda , A)}\leq  \int_{-\infty}^{\infty}F_{\beta, A}(x) H_{\Lambda , A} F_{\beta , A}(x)\ dx
\end{equation}
can be established for the excited levels of the constructed potentials by a judicious choice of the $\beta$ $N$-tuples, as each $E_{n}^{(\Lambda , A)}$ is a stationary value of the functional
\begin{equation}\label{simple.bound}
E^{(\Lambda , A)}[\Phi]=\frac{\int_{-\infty}^{\infty}\Phi(x)H_{\Lambda, A}\Phi(x) dx}{\int_{-\infty}^{\infty}\Phi(x)\Phi(x) dx}\quad .
\end{equation}
Yet, there are at most $N-1$ independent combinations $F_{\beta, A}$ and, if the overlap integrals  $\mathcal{O}_{j,k}^{(A)}$ were all to vanish, there would be exactly $N-1$ linearly independent $F_{\beta, A}$. This implies that (\ref{bounds.momma}) can only become a meaningful estimates of an expected energy only for the first $N-1$ excited states, and this only in cases when the overlap integrals can be neglected.

Let us define for each $N$-tuple $A\in\mathbb{R}^{N}$ the quantity
\begin{equation}
\vert A\vert =\max_{a_{j},a_{k}\in A}\vert a_{j}-a_{k}\vert
\end{equation}
and restrict our attention to initial potentials $V$ complying with the condition
\begin{equation}\label{1.indy.con}
\lim_{\vert A\vert\rightarrow \infty} \mathcal{O}_{j,k}^{(A)}=0 \quad \forall j,k\leq N\ .
\end{equation}
which makes the $N-1$ linear combinations $F_{\beta , A}$ ``linearly independent in the $\vert A\vert\rightarrow \infty$ limit.''
Then we get, from (\ref{alice.in.chains}) and (\ref{bounds.momma}), the result
\begin{equation}\label{big.bound}
\lim_{\vert A\vert\rightarrow \infty} E_{n}^{(\Lambda , A)}\leq E_{0}+\sum\limits_{k=1}^{N}\beta_{n; k}^{2}\Bigg(-\langle \ V\rangle+ \lim_{\vert a\vert\rightarrow\infty}\mathcal{V}_{k,k}^{(\Lambda , A)} \Bigg)\ .
\end{equation}
Furthermore, if the condition
\begin{equation}\label{2.indy.con}
\lim_{\vert A\vert\rightarrow \infty} \mathcal{V}_{n,n}^{(\Lambda , A)}=\langle V\rangle \quad \forall n\leq N
\end{equation}
is imposed on an initial potential $V$  complying with (\ref{1.indy.con}) and a  $N$-tuple $\Lambda$ complying with (\ref{convex.def}) and (\ref{convex.def.II}), then inequality (\ref{big.bound}) reduces to
\begin{equation}\label{grand.degenerate}
\lim_{\vert A\vert\rightarrow \infty} E_{n}^{(\Lambda , A)}=E_{0},\textrm{ for } 0<n< N\ .
\end{equation}
%
%
%
%

\section{A family of symmetric triple wells}
\begin{figure}[!htb]
\centering
\includegraphics[scale=1.2]{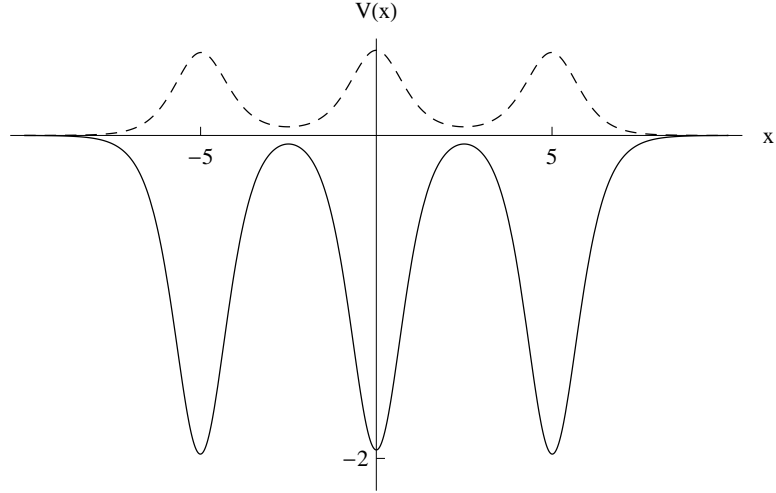}
\caption{An example of a $V_{\lambda , a}$ potential (solid curve), along with its ground eigenfunction $\Psi_{0}^{(\lambda , a)}(x)$ (dashed curve,in arbitrary units). In this case $\lambda=2/3$ and $a=5.$}
\label{fig:1GFH}
\end{figure}
Consider an initial potential
\begin{equation}\label{j-potentials}
V(x)=-2\sech ^{2} x
\end{equation}
which has a bounded spectrum consisting in a single level $E_{0}=-1$, with corresponding eigenfunction
\begin{equation}\label{j-ground}
\psi_{0}(x)=\frac{1}{\sqrt{2}}\sech  x \ ,
\end{equation}
and a continuous spectra that starts at  $E=0$. With the procedure outlined in Section 2, the following family of even triple wells:
\begin{equation}\label{3w}
V_{\lambda,a}(x)=-2\frac{[1-\lambda]\sech ^{3}x+\frac{\lambda}{2}\sech ^{3}(x+a)+\frac{\lambda}{2}\sech ^{3}(x-a)}{[1-\lambda]\sech \ x+\frac{\lambda}{2}\sech (x+a)+\frac{\lambda}{2}\sech (x-a)}
\end{equation}
can be constructed, that depends on two real parameters: $0<\lambda<1$ and $0<a$.
From Section 2 we know that each $V_{\lambda, a}$ has ground level $E_{0}=-1$ and that the ground level eigenfunction is given by
\begin{equation}\label{Ur.ground}
\Psi_{0}^{( \lambda, a)}(x)=\frac{\alpha_{\lambda ,a}}{\sqrt{2}}\Big\{[1-\lambda]\sech x+\frac{\lambda}{2}\sech (x+a)+\frac{\lambda}{2}\sech (x-a)\Big\}
\end{equation}
with a normalization constant
\begin {equation}
\alpha_{\lambda , a}=\Bigg\{[1-\lambda ]^{2}+2[1-\lambda]\lambda a\ \csch  a+\frac{\lambda^{2}}{2}(1+2a\ \csch 2a)\Bigg\}^{-1/2}
\end{equation}
that complies with the condition
\begin{equation}\label{alpha.limit}
\lim_{a\rightarrow \infty} \alpha_{\lambda ,a}=\Bigg([1-\lambda]^{2}+2\Big[\lambda/2\Big]^{2}\Bigg)^{-1/2},
\end{equation}
A typical member of the $V_{\lambda, a}$ is depicted in Figure ~\ref{fig:1GFH}, along with its ground eigenfunction.
 
For the $V_{ \lambda, a}$ family of potentials the probability density for the ground state is given by
\begin{eqnarray}
\fl\vert\Psi_{0}^{(\lambda , a)}(x)\vert^{2}=\rho_{\lambda , a}(x)= \nonumber \\
\iota_{\lambda , a}(x)+\alpha_{\lambda , a}^{2}\Big([1-\lambda]^{2}\rho (x)+\frac{\lambda^{2}}{4}\rho(x+a)+\frac{\lambda^{2}}{4}\rho(x-a) \Big)\label{spec.p.density}
\end{eqnarray}
where $\rho (x)=\vert \psi_{0}(x)\vert^{2}=\left(\sech ^{2} x\right)/{2}$ and the overlap term $\iota_{\lambda , a}$ is given by
\begin{eqnarray}
\fl \frac{\iota_{\lambda , a}(x)}{a_{\lambda , a}^{2}}=[1-\lambda ]\lambda\sech x \frac{\sech (x+a)+\sech (x-a)}{4}+ \nonumber \\
\frac{\lambda^{2}}{8}\sech(x+a)\sech(x-a)
\end{eqnarray}
\begin{figure}[!htb]
\centering
\includegraphics[scale=1.2]{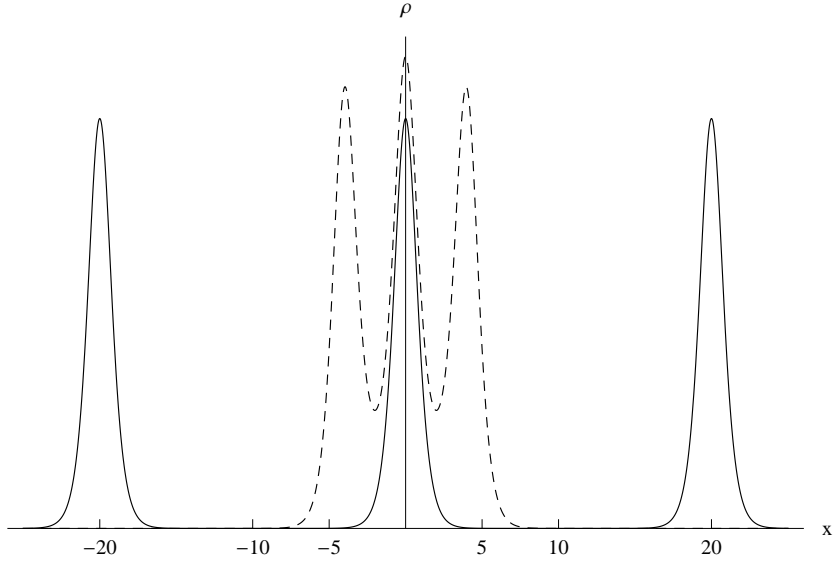}
\caption{The overlap of independent localized solutions diminishes as $a\rightarrow\infty$. Examples from the $V_{\lambda, a}$ family.  In the solid curve $\rho$ is the probability density  for $a=20$, in the dashed curve $\rho (x)$ is the probability density for $a=4$. In both cases $\lambda=2/3$. }
\label{fig:2GFH}
\end{figure}
so that for any given fixed $x\in\mathbb{R}$
\begin{equation}\label{iota.limit}
\lim_{a\rightarrow \infty}\frac{\iota_{\lambda , a}(x)/\alpha_{\lambda , a}^{2}}{[1-\lambda]^{2}\rho (x)+\frac{\lambda^{2}}{4}\rho(x+a)+\frac{\lambda^{2}}{4}\rho(x-a)}=0.
\end{equation}
Equations (\ref{alpha.limit}) and (\ref{iota.limit}) allow us to conclude that for the $V_{j, \lambda, a}$ family, the probability density for the ground state does indeed approximate to the limit
\begin{eqnarray}
\fl\rho_{\lambda , a}^{\mathrm{class}}(x)=\Bigg([1-\lambda]^{2}+2\Big[\lambda/2\Big]^{2}\Bigg)^{-1/2} \nonumber \\ %
\Bigg[\frac{[1-\lambda]^{2}}{2}\sech ^{2} x+\frac{\lambda^{2}}{8}\sech ^{2} (x+a)+\frac{\lambda^{2}}{8}\sech ^{2} (x-a)\Bigg] \label{dat.limit}
\end{eqnarray}
when $a\gg 1$. Figure ~\ref{fig:2GFH} illustrates how the overlap between solutions in different wells tends to vanish as the distances between wells increases, leading to several ``almost linearly independent solutions." 

\begin{figure}[!htb]
\centering
\includegraphics[scale=1.2]{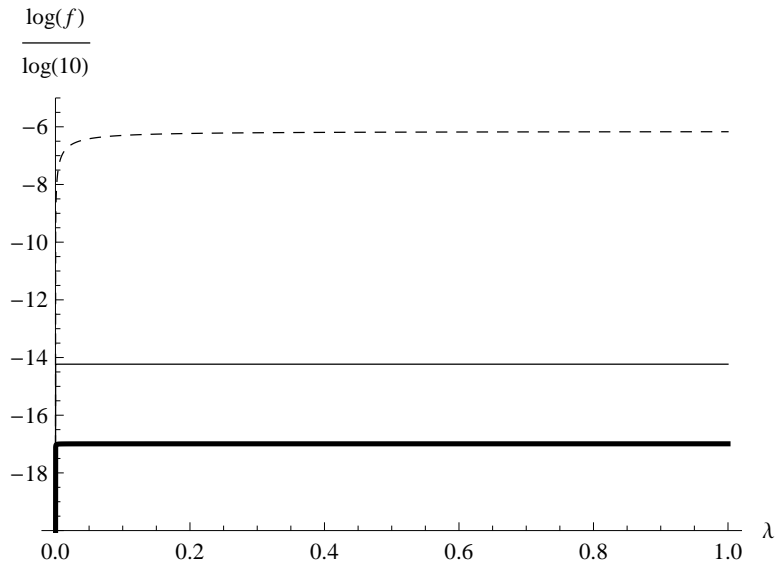}
\caption{The logarithm of function $f_{a}(\lambda)$ of (\ref{f.unction}) for $a=4$ (dashed curve), $a=7$ (solid curve) and $a=10$ (thick curve).}
\label{fig:3GFH}
\end{figure}

An upper bound can given for the first excited energy level with the use of the normalized trial function
\begin{equation}\label{Phi.def.1}
\Phi _{1}(x)=\frac{\psi_{0}(x+a)-\psi_{0}(x-a)}{\sqrt{2\Big(1-\mathcal{O}_{+,-}^{(a)}\Big)}}
\end{equation}
which, as required for a first excited eigenfunction of even potential, is odd with a single node. The overlap integral $\mathcal{O}_{+,-}^{(a)}$ appearing in (\ref{Phi.def.1}) is given by
\begin{equation}\label{specific.overlap.1}
\mathcal{O}_{+,-}^{( a)}=\int_{-\infty}^{\infty}\psi_{0}(x+a)\psi_{0}(x-a)\ dx = 2a\csch  2a
\end{equation}
Where $\csch$ stands for the hyperbolic cosecant
After some algebra, shown in the appendix, we get the following result
\begin{equation}\label{FBresult}
E_{1}\leq E_{0}+4a\frac{\frac{1-\lambda }{\lambda}\csch a+2\csch 2a}{1-2a\csch 2a}
\end{equation}
Much more important is the upper bound can be established for the second excited level, $E_{2}$, of a $V_{\lambda , a}$ potential, by using the normalized trial function
\begin{equation}\label{2.symm.bound}
\Phi_{2}(x)=\frac{[1-\lambda ]\psi_{0}(x)-\frac{\lambda}{2}\psi_{0}(x+a)-\frac{\lambda}{2}\psi_{0}(x-a)}{\sqrt{[1-\lambda]^{2}+\frac{\lambda^{2}}{2}+\frac{\lambda^{2}}{2}\mathcal{O}_{+,-}^{(a)}-2\lambda[1-\lambda]\mathcal{O}_{0,+}^{(a)}}}\ ,
\end{equation}
which is an even function with two nodes, as is required of eigenfunctions corresponding to the second excited level of an even potential.
The overlap integral $\mathcal{O}_{0,+}^{(a)}$ appearing in (\ref{2.symm.bound}) is given by
\begin{eqnarray}
\fl 
\mathcal{O}_{0,+}^{(a)}=\int_{-\infty}^{\infty}\psi_{0}(x)\psi_{0}(x+a) dx=\int_{-\infty}^{\infty}\psi_{0}(x)\psi_{0}(x-a)\ dx   \nonumber \\
=  a\csch  a \label{specific.overlap.2}.
\end{eqnarray}
After some algebra, discussed in the appendix, the bound (\ref{bounds.momma}) reduces for the this case to
 \begin{equation}\label{u.bound.2}
E_{2}-E_{0}\leq \frac{6\lambda^{2}a \csch 2a}{[1-\lambda]^{2}+\frac{\lambda^{2}}{2}+\lambda^{2}a\csch 2a-2\lambda[1-\lambda]a\csch a}\ .
\end{equation}
The right hand side of inequality (\ref{u.bound.2}), \emph{i. e.}
\begin{equation}\label{f.unction}
f_{a}(\lambda)= \frac{6\lambda^{2}a \csch 2a}{[1-\lambda]^{2}+\frac{\lambda^{2}}{2}+\lambda^{2}a\csch 2a-2\lambda[1-\lambda]a\csch a}\ ,
\end{equation}
can be shown to be an monotonically increasing function of $\lambda$ for each fixed value of $a$, which allows us to calculate a global bound for each $a$. Figure~\ref{fig:3GFH} shows that for values of $a\geq 4$ the gap $\Delta E_{0,2}=E_{2}-E_{0}$ is bounded by
\begin{equation}\label{damn.limit}
\Delta E_{0,2}<10^{-6}.
\end{equation}
As $E_{0}=-1$, with the continuous spectra starting at $E=0$, we can conclude that  for $a>4$ the three bound levels of the triple well behave as an effective three level system at $0^{\circ}\textrm{K}$.

\section{The role of $\lambda$}
\begin{figure}[!htb]
\centering
\includegraphics[scale=1.2]{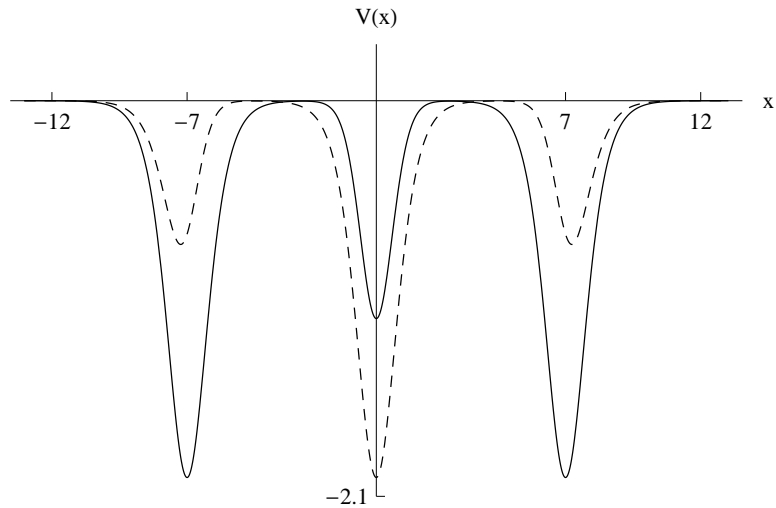}
\caption{The role of $\lambda$ on the potentials. Examples of $V_{\lambda , a}$ potentials for $a=7$. The solid curve is the case for $\lambda=0.995$, the dashed curve is the case for $\lambda=0.002$.}
\label{fig:YOFH03}
\end{figure}
\begin{figure}[!htb]
\centering
\includegraphics[scale=1.2]{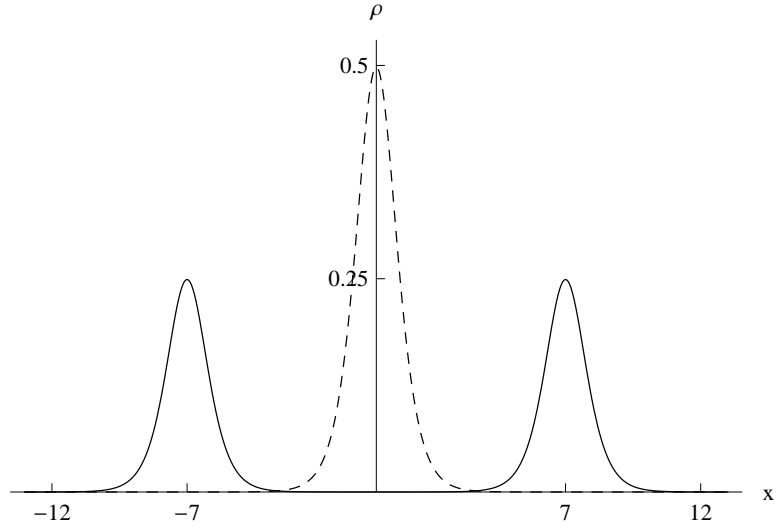}
\caption{The role of $\lambda$ on the ground probability density. The probability densities of the ground states of the potentials rendered in Fig. \ref{fig:YOFH03}. The solid curve is the case for $\lambda=0.995$, the dashed curve is the case for $\lambda=0.002$ (In both cases $a=7$).}
\label{fig:YOFH04}
\end{figure}

Graphical analysis shows that for values of $a\approx 5$  a change in the value of $\lambda$ may bring simultaneously a somewhat modest change in the shape of the potential and a complete change in the nature of the ground level state. Indeed, by changing the relative depth of the wells (Figure~\ref{fig:YOFH03}) the ground level probability density may transit from bimodal, with a peak around each of the lateral wells, to unimodal, with a single central peak (Figure~\ref{fig:YOFH04}).

\begin{figure}[!htb]
\centering
\includegraphics[scale=1.2]{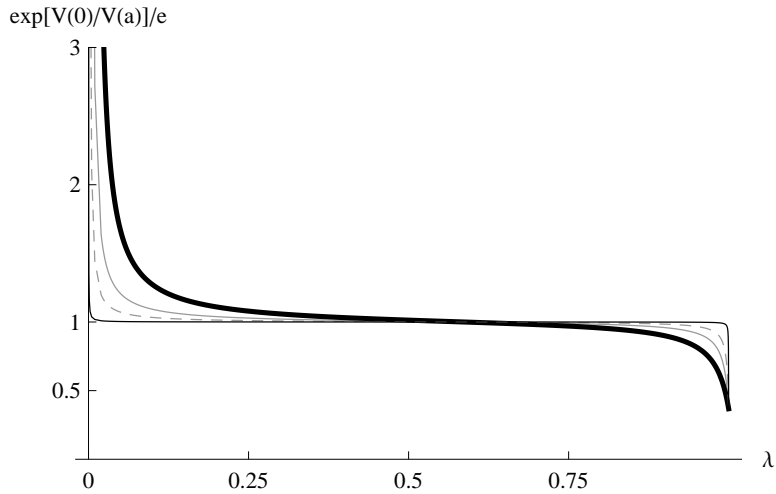}
\caption{The quotient $Q_{a}(\lambda)=V_{\lambda , a}(x=0)/V_{\lambda , a}(x=a)$ becomes virtually independent of $\lambda$ when $a\geq 10$. Thick solid curve: $a=5$, gray solid curve: $a=6$, dashed gray curve $a=7$, thin solid curve $a=10$.}
\label{fig:YOFH05}
\end{figure}

\begin{figure}[!htb]
\centering
\includegraphics[scale=1.2]{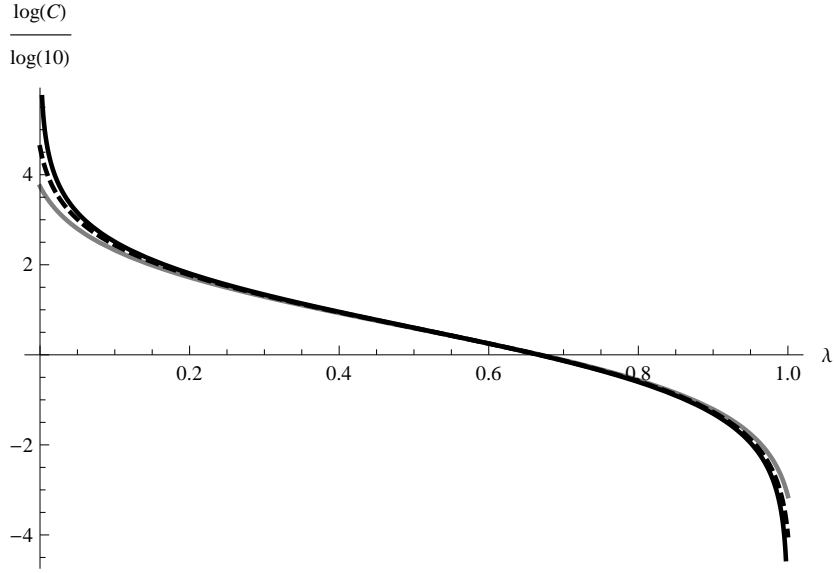}
\caption{The quotient  $C_{a}(\lambda)=\rho^{(\lambda, a)}(0)/\rho^{(\lambda, a)}(a)$ decreases by a factor of over $10^{8}$ when $\lambda$ transits from $\lambda=0$ to $\lambda=1$, for $a\geq 20$. Gray curve: $a=5$, dashed: $a=6$, black solid: $a=20$.}
\label{fig:UIAGFH72}
\end{figure}

\begin{figure}[!htb]
\centering
\includegraphics[scale=1.2]{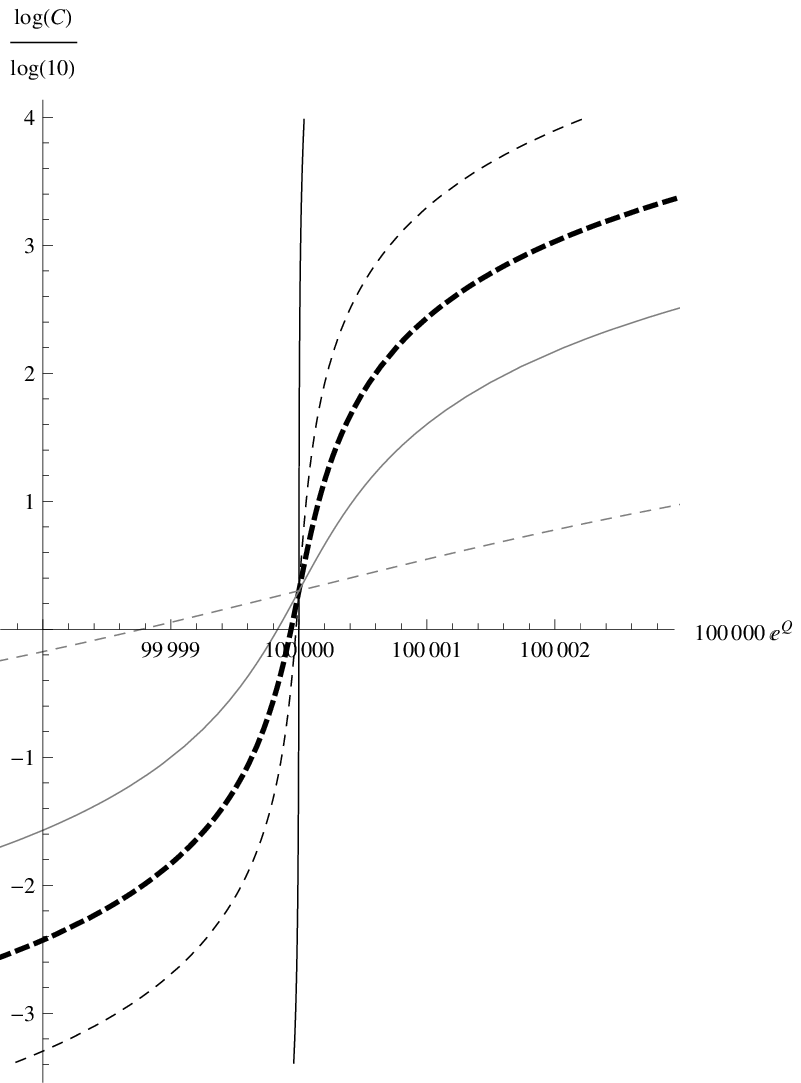}
\caption{Parametric plots showing the relationship between $e^{Q_{a}(\lambda)}$ and $\ln C_{a}(\lambda)$. Gray dashed line: $a=12$. Gray solid curve: $a=14$. Black thick dashed curve: $a=15$. Black thin dashed curve: $a=16$. Thin solid black line: $a=20$. As can be seen, the rate of change $dC_{a}/dQ_{a}$ increases without bound as  $a\rightarrow 20$. In other words, the probability density becomes exquisitely sensitive to changes $V_{\lambda , a}\rightarrow V_{\lambda+\delta, a}$ as $a$ approaches $a=20$ from below.}
\label{fig:UIAGFH47}
\end{figure}

Moreover, as $a\rightarrow\infty$ the relative depth of the wells of the potential, which can be approximated by the quotient
\begin{equation}
Q_{a}(\lambda)=\frac{V_{\lambda , a}(0)}{V_{\lambda , a}(a)},
\end{equation}
 becomes practically independent of the value of $\lambda$ in the interval $0<\lambda<1$ (Figure~\ref{fig:YOFH05},) with all three local minima converging to a single value $V_{L}$. On the other hand, the quotient of the ground probability density peaks can be approximated by the quotient
\begin{equation}
C_{a}(\lambda)=\frac{\rho^{\lambda , a}(0)}{\rho^{\lambda , a}(a)},
\end{equation}
which varies from over $10^{4}$ for $\lambda=0$, to under $10^{-4}$ for $\lambda=1$, for $a\geq 20$ (Figure~\ref{fig:UIAGFH72}). The combination of the two behaviours gives rise to an exquisite sensitivity shown by the ground probability distribution towards the shape of  the potential $V_{\lambda , a}$ as the distance $a$ approaches the value $a=20$ from below (Figure~\ref{fig:UIAGFH47}.)
\section{Length and times scales}
Up to this point in the present paper we have been working only with adimensional versions of the Schroedinger equation. In order to obtain, starting from an adimensional operator such as (\ref{initial.H}), a Hamiltonian $\mathfrak{H}$ with the correct dimensions, one has to put by hand the mass $m>0$ of an actual system to be described, and its typical length $L>0$. In this way one obtains:
\begin{equation}\label{adim.dim}
\mathfrak{H}=\frac{\hbar^{2}}{2mL^{2}}H
\end{equation}
This dimensionally correct Hamiltonian is now written as
\begin{equation}\label{dim.H}
\mathfrak{H}=-\frac{\hbar^{2}}{2m}\frac{d^{2}}{d\xi^{2}}+U(\xi)
\end{equation}
where $\xi=Lx$ and $U(\xi)$ is related to the adimensional potential $V(x)$ of (\ref{initial.H}) through
\begin{equation}\label{dim.pot}
U(\xi)=\frac{\hbar^{2}}{2mL^{2}}V(\xi /L).
\end{equation}
Analogous expressions can be written for the elements $\mathfrak{H}_{\lambda , a}$ of a family of dimensionally correct Hamiltonians, and for their dimensionally correct potentials $U_{\lambda , a}(\xi)$.  Then, dimensionally correct  wave functions are obtained according with the law
\begin{equation}
\psi (x)\rightarrow \tilde{\psi}(\xi)=L^{-1/2}\psi(\xi/L)
\end{equation}
so that the values of the overlap integrals (and thus, the values of the $\alpha_{\lambda , a}$ normalization factors) remain independent of $L$:
\begin{equation}
\int \psi (x+a_{j})\psi(x+a_{k})dx =\int \tilde{\psi}(\xi+La_{j}) \tilde{\psi}(\xi+La_{k})d\xi\ .
 \end{equation}
All distances pertaining to the system are just multiples of $L$ times the corresponding adimensional parameter, including the minimum  distance $L\vert A\vert$ (corresponding, loosely, to the minimum distance between probability peaks of states $\tilde{\psi}_{0}(\xi+La_{j})$, or the alternatively to the minimum distance between the minima of the dimensionally correct potential $U(\xi)$) and the dimensionally correct dispersion $D_{\xi}$,
\begin{equation}
D_{\xi}=\sqrt{\int_{-\infty}^{\infty}\vert\tilde{\psi}(\xi)\vert^{2}\xi^{2}dx-\Bigg[\int_{-\infty}^{\infty}\vert\tilde{\psi}(\xi)\vert^{2}\xi d\xi\Bigg]^{2}}=LD_{x} .
\end{equation}
On the other hand, all energies $\mathfrak{E}$ pertaining to the system scale as
\begin{equation}
E\rightarrow  \mathfrak{E}=\frac{\hbar^{2}}{2mL^{2}}E
\end{equation}
so that the dimensionally correct versions of all relevant frequencies,$\Omega$, are obtained from their adimensional counterparts, $\omega$, through:
\begin{equation}
\omega\rightarrow \Omega=\frac{\hbar}{2mL^{2}}\omega .
\end{equation}
The value of the disspersion in the position for the localized eigenfunction $\psi_{0}(x)$ of (\ref{j-ground}) can be estimated as
\begin{equation}\label{last1}
D_{x}=2.34
\end{equation}
through numerical integration. Taking $D_{\xi}\approx 5\times 10^{-8}$ m from the typical value for the well width in semiconductor heterostructures, one obtains the typical lenght $L$ consistent with our model as
\begin {equation}
L=D_{\xi}/D_{x}\approx 2\times 10^{-8} \textrm{ m. }
\end{equation}
Setting $m$ as the electron rest mass, \emph{i. e.} $m=m_{e}=1.7\times 10^{-27}$ kg, thus gives
\begin {equation}
\frac{\hbar}{2mL^{2}}\approx 6\times 10^{7}s^{-1}.
\end{equation}
From this and equation (\ref{damn.limit}) we can conclude that even for a modest value of $a=4$ the periods associated with the low-lying three level system of potential $V_{\lambda , a}$  in a mesoscopic heterostructure would be of the order of one  hundredth of a second or bigger,
\begin{equation}\label {latest0}
\Omega\leq\frac{\hbar}{2mL^{2}}\Delta E_{0,2} \lesssim 2\times 10^{-1}s^{-2},
\end{equation}
which is some two orders of magnitude above the maximum period that can be observed before dissipation destroys coherent oscillations\cite{Leggett80} ($\tau\approx 100\mu$s.)

\section{Discussion}
We have reasons to believe that the sensitivity-growing-with-distance effect is not a pathology, the outcome of a poorly chosen example of a potential. For one thing, the results here laid out are easily generalized to a wider family of triple wells, \emph{viz} those obtained by taking as initial $V$ a member of the hyperbolic Rosen-Morse or hyperbolic Scarf families of potentials \cite{Gangopadhyaya}. We have simply chosen an example that lends itself to be exposed in a few pages. Also, we have restricted ourselves to models with exactly solvable ground levels, but it may be possible that the procedure of Section~2 can be adapted in order to include square wells, which would foreseeable agree with the results presently under discussion. Moreover, our model reproduces well established features of the behaviour of submicron semiconductor heterostructures, such  as the quantum conﬁnement effect, which predicts a band gap decrease with increasing system size,\cite{MedeirosRibeiroetal, Chang99, Jiang2010, Arivazhagan2013} as attested by equation (\ref{u.bound.2}) and Figure~\ref{fig:3GFH}. Not the least, Figure~\ref{fig:UIAGFH47} can also be interpreted as implying that for $a\geq 20$ the ground level becomes effectively degenerate (as any possible linear combination is then an acceptable stationary solution with only an infinitesimal change in the potential.)

Even if the gap between the frequencies predicted by equation (\ref{latest0}) and those experimentally observable is narrowed, either by picking a more suitable model or by any future technological development, the breach is too wide as to make it dubious that it can ever be completely filled. If this gap is truly unsormountable, \emph{i. e. } if there exists a physical (as opposed to merely technological) upper limit for the periods of observable coherent oscillations, then the model described in the preceeding pages would be that of a system for which the precise quantum description can neither be proven nor disproven, but simply results irrelevant. Instead, the semi-classical description (including a degenerate ground level  with independent localized solutions) would be as complete as possible, and without incurring in unfalsifiable predictions. That is, the system would be essentially semi-classical. So this example may be of interest when testing the limits of quantum theory in the mesoscopic realm.\cite{Leggett2002}

On the other hand, if the sensitivity-growing-with-distance effect described in this paper can be observed, even by indirect means, that may be of relevance in applied physics. In any case we believe that the question merits further attention.

\ack{
The support of CONACYT of duly acknowledged . RMV also  acknowledges the financial support of FICSAC (UIA) and the sabbatical leave program of UACM, as well as the warm hospitality of \emph{Departamento de F\'{i}sica y Matem\'{a}ticas} (UIA.)
}
%
%
%
%
\appendix
\section{Appendix}
Given the symmetry of the $V_{\lambda , a}$ potentials, expression  (\ref{simple.bound}) reduces to
\begin{equation}\label{App.1}
E_{1}\leq E_{0}+\frac{\mathcal{V}_{+,+}^{(\lambda , a)}-\langle V\rangle-\mathcal{V}_{+,-}^{(\lambda , a)}+\mathcal{U}_{+,-}^{(a)}}{1-\mathcal{O}_{+,-}^{(a)}},
\end{equation}
for the case under discussion, with the overlap integral $\mathcal{O}_{+,-}^{(a)}$ given in (\ref{specific.overlap.1}), the integral $\langle V  \rangle$ as  in (\ref{expect.v})  and the $\mathcal{V}$ and $\mathcal{U}$ integrals being given by
\begin{eqnarray}
\mathcal{V}_{+,+}^{(\lambda , a)} &=& \int_{-\infty}^{\infty}\psi_{0}(x+a)V_{\lambda , a}(x)\psi_{0}(x+a)\ dx \nonumber \\
 &=& \int_{-\infty}^{\infty}\psi_{0}(x-a)V_{\lambda , a}(x)\psi_{0}(x-a)\ dx\ \\
\mathcal{V}_{+,-}^{(\lambda , a)} &=& \int_{-\infty}^{\infty}\psi_{0}(x+a)V_{\lambda , a}(x)\psi_{0}(x-a)\ dx \\
\mathcal{U}_{+,-}^{(a)} &=& \int_{-\infty}^{\infty}\psi_{0}(x+a)V(x-a)\psi_{0}(x-a)\ dx \nonumber \\
& =&\int_{-\infty}^{\infty}\psi_{0}(x-a)V(x+a)\psi_{0}(x+a)\ dx .
\end{eqnarray}
As the initial potential $V=-2\sech ^{2} x$ has as least upper bound $V_{U}=0$ and as greatest lower bound $V_{L}=-2$ then the relation
\begin{equation}\label{App.2}
E_{1}\leq E_{0}+\frac{\mathcal{V}_{+,+}^{(\lambda , a)}-\langle V\rangle}{1-\mathcal{O}_{+,-}^{(a)}}+\frac{2\mathcal{O}_{+,-}^{(a)}}{1-\mathcal{O}_{+,-}^{(a)}}
\end{equation}
follows from (\ref{App.1}).

The expression  $\mathcal{V}_{+,+}^{(\lambda , a)}-\langle V\rangle$ appearing in (\ref{App.2}) can be written in following alternative manner
\begin{eqnarray}
\fl
\mathcal{V}_{+,+}^{(\lambda , a)}-\langle V\rangle = \int_{-\infty}^{\infty}\psi_{0}(x+a)\frac{\sum [V(x+a_{n})-V(x+a)]\lambda_{n}\psi_{0}(x+a_{n})}{\sum \lambda_{n}\psi_{0}(x+a_{n})}\psi_{0}(x+a)\ dx\ \nonumber \\
 = \frac{1-\lambda}{\lambda/2}\int_{-\infty}^{\infty}\psi_{0}(x+a)[V(x)-V(x+a)]\psi_{0}(x) \frac{\frac{\lambda}{2}\psi_{0}(x+a)}{\sum \lambda_{n}\psi_{0}(x+a_{n})}\ dx \nonumber \\
 +  \int_{-\infty}^{\infty}\psi_{0}(x+a)[V(x-a)-V(x+a)]\psi_{0}(x-a) \frac{\frac{\lambda}{2}\psi_{0}(x+a)}{\sum \lambda_{n}\psi_{0}(x+a_{n})}\ dx \label{App.3}
\end{eqnarray}
As each term in the sum is strictly positive, then
\begin{equation}\label{App.4}
\frac{\frac{\lambda}{2}\psi_{0}(x+a)}{\sum \lambda_{n}\psi_{0}(x+a_{n})}<1
\end{equation}
and thus, from (\ref{App.3}) and (\ref{App.4}) we get the inequality
\begin{equation}\label{App.5}
\mathcal{V}_{+,+}^{(\lambda , a)}-\langle V\rangle \leq  2\frac{1-\lambda}{\lambda/2}\mathcal{O}_{0,+}^{(\lambda ,a)}+2\mathcal{O}_{+.-}^{(\lambda ,a)}
\end{equation}
where the overlap integral $\mathcal{O}_{0,+}^{(a)}$ is as given in (\ref{specific.overlap.2}). Plugging (\ref{App.5}) in (\ref{App.2}) we obtain
\begin{equation}\label{App.6}
E_{1}\leq E_{0}+4\frac{\frac{1-\lambda}{\lambda}\mathcal{O}_{0,+}^{(\lambda ,a)}+\mathcal{O}_{+.-}^{(\lambda ,a)}}{1-\mathcal{O}_{+,-}^{(a)}}
\end{equation}
and, finally, plugging (\ref{specific.overlap.1}) and (\ref{specific.overlap.2}) in (\ref{App.6}) gives us (\ref{FBresult}), which is the first of the results we set out to prove in this appendix.
As for the bound for second excited level, the reflection symmetry of the $V_{\lambda, a}$ potentials again reduces in an important way the number integrals that must be evaluated in order to calculate the bound (\ref{u.bound.2}). The result can be written as
\begin{eqnarray}
\fl E_{2}\leq E_{0}\ -\langle \ V\rangle\big(\beta_{0}^{2} + 2\beta_{+}^{2}\big)  + \beta_{0}^{2}\mathcal{V}_{0,0}^{(\lambda , a)}  +  2\beta_{+}^{2}\mathcal{V}_{+,+}^{(\lambda , a)}\nonumber\\
  +  4\beta_{0}\beta_{+}(\mathcal{V}_{0,+}^{(\lambda , a)}-\mathcal{U}_{0,+}^{(a)}) +  4\beta_{+}^{2}(\mathcal{V}_{-,+}^{(\lambda , a)} - \mathcal{U}_{-,+}^{(a)})\label{long.2.bound}
\end{eqnarray}
with the use of the linear coefficients $\beta_{0}$, and $\beta_{+}=\beta_{-}$ given in (\ref{2.symm.bound}), that is
\begin{equation}\label{App.7}
\beta_{0}=\frac{1-\lambda}{\sqrt{[1-\lambda]^{2}+\frac{\lambda^{2}}{2}+\frac{\lambda^{2}}{2}\mathcal{O}_{+,-}^{(a)}-2\lambda[1-\lambda]\mathcal{O}_{0,+}^{(a)}}}
\end{equation}
and
\begin{equation}\label{App.8}
\beta_{+}=\frac{-\lambda/2}{\sqrt{[1-\lambda]^{2}+\frac{\lambda^{2}}{2}+\frac{\lambda^{2}}{2}\mathcal{O}_{+,-}^{(a)}-2\lambda[1-\lambda]\mathcal{O}_{0,+}^{(a)}}}\ .
\end{equation}
The definitions of the four $\mathcal{V}$ integrals and the two $\mathcal{U}$ integrals appearing in (\ref{long.2.bound}) are completely analogous to definitions (\ref{nother.bee}) and (\ref{bumble.bee}). Integral $\langle V\rangle$ is as defined in (\ref{expect.v}).

As the initial potential complies with $-2\leq V(x)< 0$ for all $x\in\mathbb{R}$,  then inequality
\begin{eqnarray}
\fl E_{2}\leq E_{0}\ -\langle \ V\rangle\big(\beta_{0}^{2}+2\beta_{+}^{2}\big)  +  \beta_{0}^{2}\mathcal{V}_{0,0}^{(\lambda , a)}  +  2\beta_{+}^{2}\mathcal{V}_{+,+}^{(\lambda , a)} 
+  8(-\beta_{0}\beta_{+}\mathcal{O}_{0,+}^{(a)} + \beta_{+}^{2}\mathcal{O}_{-,+}^{(a)})\label{short.bound}
\end{eqnarray}
follows from (\ref{long.2.bound}).

The inequality
\begin{equation}\label{App.9}
\mathcal{V}_{0,0}^{(\lambda , a)}-\langle V\rangle < \frac{2\lambda}{1-\lambda}\mathcal{O}_{0,+}^{(a)}
\end{equation}
is obtained in a way similar to was what done for (\ref{App.5}), and from  (\ref{App.5}) (\ref{short.bound}) and (\ref{App.9}) we get
\begin{eqnarray}
\fl E_{2}\leq E_{0}  + \beta_{0}^{2}\frac{2\lambda}{1-\lambda}\mathcal{O}_{0,+}^{(a)} + 4\beta_{+}^{2}\Big(2\frac{1-\lambda}{\lambda}\mathcal{O}_{0,+}^{(a)}+\mathcal{O}_{+.-}^{(a)}\Big) \nonumber \\
+ 8\Big(-\beta_{0}\beta_{+}\mathcal{O}_{0,+}^{(a)}+\beta_{+}^{2}\mathcal{O}_{-,+}^{(a)} \Big)\label{App.10}.
\end{eqnarray}
Substituting (\ref{App.7}) and (\ref{App.8}) in (\ref{App.10}) gives us
 \begin{equation}\label{App.11}
E_{2}\leq E_{0}\ +\frac{3\lambda^{2}\mathcal{O}_{+,-}^{(a)}}{[1-\lambda]^{2}+\frac{\lambda^{2}}{2}+\frac{\lambda^{2}}{2}\mathcal{O}_{+,-}^{(a)}-2\lambda[1-\lambda]\mathcal{O}_{0,+}^{(a)}}\ ,
\end{equation}
and substituting (\ref{specific.overlap.1}) and (\ref{specific.overlap.2}) in (\ref{App.11}) finally gives us (\ref{u.bound.2}) which is the second and last result we set put to prove in this appendix.



\providecommand{\newblock}{}

\end{document}